# COMMUTATIVITY OF FIRST-ORDER DISCRETE-TIME LINEAR TIME-VARYING SYSTEMS

Mehmet Emir KOKSAL

*Department of Mathematics, Ondokuz Mayis University, 55139 Atakum, Samsun, Turkey*

*emir_koksal@hotmail.com*

**Abstract:** After introducing the concept of commutativity for continuous-time linear time-varying systems, the related literature and the results obtained so far are presented. For a simple introduction of the commutativity of discrete-time linear time-varying systems, the problem is formulated for first-order systems. Finally, explicit necessary and sufficient conditions for the commutativity of first-order discrete-time linear time-varying systems are derived, and their advantageous use in digital system design is illustrated; which are the main objectives of the paper. The results are verified by examples which include an application in amplitude modulation for digital telecommunication.

## I. INTRODUCTION

Many engineering systems are in consecutive form structurally so that the output of one section is the input of the next. Especially in electrical and electronics engineering, many circuits such as electrical filters are designed in this form which is known to be cascade or chain connection [1-5]. The order of the connection of subsystems in the chain may be compatible with the physical system model or may depend on the special synthesis method in a design problem. In cases where sensitivity, stability, linearity, noise disturbance and robustness effects are of concern, a certain ordering may be preferable over the other though they both yield the same main functioning [6, 7]. Hence, the commutativity becomes apparently important for such cases in practice.

In particular, cascade connection of electrical circuits is important at least as series and parallel connections in electrical and electronics engineering [8, 9]. Furthermore, time-varying linear circuits are the corner stones of modulating circuits in modern communication theory [10, 11]. Hence, the commutativity of time-varying systems is not only important from the theoretical point of view, but equally attractive for electronic engineering applications.

There are many advantages of using digital systems comparing to the analog systems. For example, digital systems can be tuned by software to achieve different applications without necessitating any changes in the hardware. In fact, sophisticated and highly cost hardware is eliminated by using digital technology due to grate advances in IC technology where now highly grate number of components can be placed in a given chip are, which helps further cost reduction. Further, many processing techniques have been developed for digital signals so that high speed of operation is possible even for very fast online operations. Another advantage of digital systems is their reliability due to possibility of using error correcting codes. They uses Boolean algebra and other digital techniques which simplify the design methods as compared to analog system design. Moreover digital systems are less affected to environmental conditions such as humidity, temperature, noise and this result with higher reproducibility when compared with analog systems.

Digital communication is one of the widest application area of digital systems. It is based on coding and encoding signals so digital devices used are mostly similar. Therefore, digital communication carries all general advantages of digital systems such as more immunity to external disturbances, cheapness, simple design, tenability, compactness, error detection and correction, high speed. It provides added security to information signal which can be saved and retrieved easily and this is not always possible in analog systems.

Due to the above-mentioned advantages, the modern communication technology is running to digital side from its analog state and it is now in the foreground [12, 13]. On the other hand, the modern technology is sliding rapidly towards the digital area from its classical analog state and digital communication is now in the foreground [12, 13]. Digital systems are modelled as discrete-time systems. In the last decade, there is a great number of papers published in the famous journals, on the theory and applications of discrete-time linear time-varying systems [14-27]. Therefore, it is very important to investigate the commutativity property for time-varying discrete-time systems as well. The paper emphasizes this importance and is aimed to be the first attempt to carry the theory developed for commutativity of analog systems [6] into discrete-time domain. One possible benefit of commutativity for discrete-time systems is also demonstrated by an example.

Consider the time-varying linear systems $A$ and $B$ described by the $n$-th order and $m$-th order differential equations

$$A: \quad \sum_{i=0}^{n} a_i(t) \frac{d^i}{dt^i} y_A(t) = x_A(t), a_n(t) \neq 0, \tag{1a}$$

$$B: \quad \sum_{i=0}^{m} b_i(t) \frac{d^i}{dt^i} y_B(t) = x_B(t), b_m(t) \neq 0 \tag{1b}$$

for $t \geq t_0$, where $a_i(\bullet)$ and $b_i(\bullet)$ are piecewise continuous functions from $R$ into $R$; $x_A$ ($x_B$) and $y_A$ ($y_B$) are the input and output of the system $A$ ($B$). It is assumed that $x_A(\bullet)$ and $x_B(\bullet): R \to R$ are also piecewise continuous. $A$ is a system of order $n \geq 1$, the order of $B$ is $m \leq n$; hence, $a_n(t) \not\equiv 0, b_m(t) \not\equiv 0$. The initial time $t_0$ could be assumed 0 without spoiling the generality since the transformation $t \to t - t_0$ achieves this property. It is well-known that with the mentioned input and coefficient function spaces both systems have unique solutions for any given set of initial conditions [28].

The cascade connection $AB$ of the systems $A$ and $B$ is defined as the combined system obtained by feeding the input $x_B$ of $B$ by the output $y_A$ of $A$, that is $y_A = x_B$; hence for $AB$, the input is $x = x_A$ and the output is $y = y_B$. Therefore, the system $AB$ is described by the $(n + m)$-th order differential equation

$$AB: \quad \sum_{k=0}^{n} a_k(t) \sum_{i=0}^{m} \frac{d^k}{dt^k} \left[ b_i(t) \frac{d^i}{dt^i} y(t) \right] = x(t). \tag{2a}$$

For the cascade connection of $BA$, a similar approach with $y_B = x_A, x = x_B, y = y_A$ yields

$$BA: \quad \sum_{i=0}^{m} b_i(t) \sum_{k=0}^{n} \frac{d^i}{dt^i} \left[ a_k(t) \frac{d^k}{dt^k} y(t) \right] = x(t). \tag{2b}$$

The systems $A$ and $B$ are said to be (zero-state) commutative if $AB$ and $BA$ yield the same solution $y(t)$ for any input $x(t)$; that is they have the same input-output pairs.

The first paper about the commutativity of continuous-time linear time-varying systems is due to E. Marshall and appeared in 1977 [29]. In that paper and in all that are cited thereafter, the commutativity of continuous-time linear time-varying systems are investigated on the base of the above mentioned definition. E. Marshall has shown the necessary and sufficient conditions for the commutativity of first-order systems and proven that for the commutativity of two linear systems, it is required that either both systems are time-invariant (scalar constant gain systems are excluded) or both systems are time-varying. After Marshall's work, there had been several publications confined to special cases [30-33] until the first exhaustive study of commutativity of continuous-time linear time-varying systems appeared in 1988 [34].

In [29-31], the commutativity results for second-order systems are presented with the contribution of S.V. Saleh [32]. Explicit commutativity conditions for the third and fourth order systems are derived in [33]. The general commutativity conditions for linear time-varying systems presented in [34] not only covered all the previous works as special cases but it also forecasted open questions on commutativity conditions with non-zero initial values, role of the performance concerning sensitivity, disturbance and noise, which are all important from engineering point of view. Later, the concept of commutativity has not been studied about 20 years until the work [6] in 2011. This work is not only a tutorial paper on the commutativity of linear time-varying systems covering the results of [34] and those of published thereafter, it also includes the explicit conditions for the commutativity of fifth-order linear time-varying systems.

Although the commutativity of continuous-time linear time-varying systems have been studied in many papers scattered in the literature on time since 1978, the most of the work on this area can be found in a single reference [6].

On the other hand, literature on commutativity of discrete-time linear time-varying systems hardly exist, and as far as the author knowledge these has appeared a single reference in 2015 [7]. It is true that continuous-time linear time-varying systems and discrete-time linear time-varying systems are quite different in nature; the first is modelled by differential equation and the second by difference equation. The theory of these two different area is hardly similar; for example Laplace transform, Green functions are some of the tools for differential equations whiles z-transform is used for studying the discrete-time linear time-varying systems.

In that reference, after introducing the commutativity concept for discrete-time linear time-varying systems, some open questions have been stated; such as

  i. *Is a discrete time-varying system commutative with its feedback control version?*
 ii. *Are the relation between the coefficients of commutative pairs similar to those of the analog systems that is, can they be related by a matrix equation?*

iii. *Can the commutativity property be used to design more robust and less sensitive discrete time-varying systems?*

iv. *Is commutativity property an entity for the system itself or does it also depend on the specific input applied?*

v. *What are the additional conditions for commutativity under nonzero initial conditions?*

By considering the second-order discrete-time linear time-varying systems only, some of the above questions (for example, iii, iv, v) have been answered in [7]. However, the first and second question have not been answered. In fact, considering the commutativity of second-order discrete-time linear time-varying systems, comparing the results for analog systems in [6] and discrete-time systems in [7], it is seen that the commutativity conditions for analog and digital systems are completely in different forms, for example a matrix equation for the coefficients of the commutative pairs cannot be obtained in the case of discrete-time systems.

The main objective of the present paper is focused on the commutativity of first-order discrete-time linear time-varying systems. For such systems, it is shown that *i.* the feedback conjugate of a first-order discrete-time linear time-varying system is always commutative with the system itself, and *ii.* The relation between the commutative pairs is expressed by a matrix equation.

It is satisfied with the introductory knowledge given in this section which will be closed by describing the content of the rest of the paper. Section 2 is devoted to the equivalence of two discrete-time linear time-varying systems. Section 3 defines and formulates the commutativity problem for the discrete-time linear time-varying systems. In Section 4, the explicit commutativity conditions for the first-order discrete-time linear time-varying systems are derived. Section 5 includes some examples validating the results of the previous section; one possible merit of commutativity is also included for discrete-time linear time-varying systems. Finally, the paper finishes with Section 5 that deals with conclusions and the suggested future work.

## II. EQUIVALENCE OF DISCRETE-TIME LINEAR SYSTEMS

Before studying commutativity, some preliminaries concerning definitions of equivalence and the related lemmas about the discrete-time linear time-varying systems are considered. Let such a system of order $n$ be described by the difference equation

$$\sum_{i=0}^{n} a_i(k) y(k+i) = x(k), \ a_n(k) \neq 0: k = 0,1,2,\cdots, \quad (3a)$$

where the coefficients $a_i(\bullet)$ and the input $x(\bullet)$ are bounded functions from the discrete-time space $Z^+ = \{0,1,2,\cdots\}$ to $R$, which is denoted by $B[Z^+]$. Let the initial conditions be represented by the initial state vector

$$[y_0 \quad y_1 \quad y_2 \quad \cdots \quad y_{n-1}] \in R^n; \ y(k) = y_k, k = 0,1,2,\cdots, n-1. \tag{3b}$$

It is obviously true that the solution of the system (3) is uniquely obtained by successive applications of the formula

$$y(n+k) = \frac{1}{a_n(k)}\left[x(k) - \sum_{i=0}^{n-1} a_i(k)y(k+i)\right], \forall k \geq 0. \tag{4}$$

Hence, the condition $a_n(k) \neq 0, \forall k \geq 0$ is a sufficient condition for the existence and uniqueness of the solution $y(n+k), \forall k \geq 0$ for any given set of initial conditions $y_0, y_1, \cdots, y_{n-1}$ and the input sequence $x(0), x(1), \cdots$; moreover $y(n+k)$ can be set to any desired value by a proper choice of $x(k)$, $\forall k \geq 0$. This condition is also necessary for the existence of a unique solution for the mentioned initial conditions and the input sequence. Because, otherwise any $m \geq 0$ for which $a_n(m) = 0$ will impose $a$ restriction on $x(m)$ depending on its previous values and/or the initial conditions.

For the definition of the commutativity under zero initial conditions, zero-state equivalence of two linear discrete-time systems of the same order is defined.

**Definition 1:** Two discrete-time linear time-varying systems of the same order $n$ described by the difference equations of type 3 are said to be zero-state equivalent if they produce the same solution for all $k \geq 0$ for any input $x \in [Z^+]$ when their initial states are zero. Note that zero-state equivalent systems have the same input-output pairs under relaxed conditions.

For the formulation of the conditions of the commutativity, the following lemma is needed:

**Lemma 1:** For the zero-state equivalence of two systems of the same type, (3) and (5)

$$\sum_{i=0}^{n} \bar{a}_i(k)\bar{y}(k+i) = \bar{x}(k), \ \bar{a}_n(k) \neq 0; k = 0,1,2,\cdots, \tag{5a}$$

$$[\bar{y}_0 \quad \bar{y}_1 \quad \bar{y}_2 \quad \cdots \quad \bar{y}_{n-1}] \in R^n \colon \bar{y}(h) = \bar{y}_h, h = 0,1,2,\cdots, n-1, \tag{5b}$$

it is necessary and sufficient that

$$\bar{a}_i(k) = a_i(k), \ \forall k \geq n-i, \ i = 0,1,2,\cdots, n. \tag{6}$$

The proof follows directly from Eq. (4) and similar equation written for the solution of (5) by considering the zero-states $y_k = \bar{y}_k = 0$ for $k = 0,1,2,\cdots,n-1$ and requiring identical solutions $y(k) = \bar{y}(k)$ for all $k \geq n$ for arbitrary equal inputs $x(k) = \bar{x}(k)$.

**Proof of Lemma 1:** Solution of (5) can be written by a similar equation to (4), that is

$$\bar{y}(n+k) = \frac{1}{\bar{a}_n(k)}\left[\bar{x}(k) - \sum_{i=0}^{n-1} \bar{a}_i(k)\bar{y}(k+i)\right]; \forall k \geq 0. \qquad (7)$$

Consider the sequence of arbitrary input values $\bar{x}(k) = x(k), \forall k \geq 0$ for the systems (3) and (5). Since for the zero-state response the initial values $y(i) = \bar{y}(i)$ are all zero for $i = 0,1,2,\cdots,n-1$, for $k = 0$, Eqs. (4) and (7) yield $y(n) = \frac{1}{a_n(0)}x(0), \bar{y}(n) = \frac{1}{\bar{a}_n(0)}x(0)$; respectively; the zero-state equivalence requires $\bar{y}(n) = y(n)$, that is

$$\bar{y}(n) - y(n) = \left[\frac{1}{\bar{a}_n(0)} - \frac{1}{a_n(0)}\right]x(0) = 0$$

For all arbitrary $x(0)$. This is satisfied if and only the coefficient of $x(0)$ is zero, that is

$$\bar{a}_n(0) = a_n(0). \qquad (8a)$$

Furthermore $\bar{y}(n) = y(n)$ can be set to any value since $x(0)$ is arbitrary.

Having $\bar{y}(i) = y(i)$ for $i = 0,1,\cdots,n$; whilst $\bar{y}(i) = y(i) = 0$ for $i = 0,1,\cdots,n-1$, and $\bar{y}(n) = y(n)$ can be set to any arbitrary value by $x(0)$, now consider the solutions (4) and (7) for $k = 1$.

$$\bar{y}(n+k) = \frac{1}{\bar{a}_n(1)}\left[\bar{x}(1) - \sum_{i=0}^{n-1} \bar{a}_i(1)\bar{y}(1+i)\right]$$

$$= \frac{1}{\bar{a}_n(1)}\left[x(1) - \sum_{i=n-1}^{n-1} \bar{a}_i(1)y(1+i)\right] = \frac{1}{\bar{a}_n(1)}x(1) - \frac{\bar{a}_{n-1}(1)}{\bar{a}_n(1)}y(n),$$

$$y(n+k) = \frac{1}{a_n(1)}\left[x(1) - \sum_{i=0}^{n-1} a_i(1)y(1+i)\right]$$

$$= \frac{1}{a_n(1)}\left[x(1) - \sum_{i=n-1}^{n-1} a_i(1)y(1+i)\right] = \frac{1}{a_n(1)}x(1) - \frac{a_{n-1}(1)}{a_n(1)}y(n).$$

The zero-state equivalence requires $\bar{y}(n+1) = y(n+1)$, that is

$$\bar{y}(n+1) - y(n+1) = \left[\frac{1}{\bar{a}_n(1)} - \frac{1}{a_n(1)}\right]x(1) - \left[\frac{\bar{a}_{n-1}(1)}{\bar{a}_n(1)} - \frac{a_{n-1}(1)}{a_n(1)}\right]y(n) = 0.$$

This equation is valid for all arbitrarily chosen $x(1)$ and arbitrary values $y(n)$ set by $x(0)$. Therefore, it is satisfied if and only if the coefficients of $x(1)$ and $y(n)$ are zero. This requires

$$\bar{a}_n(1) = a_n(1), \bar{a}_{n-1}(1) = a_{n-1}(1). \tag{8b}$$

Further, $\bar{y}(n+1) = y(n+1)$ can be set to any value independent from their previous values since $x(1)$ is arbitrary.

Having $\bar{y}(i) = y(i)$ for $i = 0,1,\cdots,n, n-1$, whilst $\bar{y}(i) = y(i) = 0$ for $i = 0,1,\cdots,n-1$, and $\bar{y}(n) = y(n)$, $\bar{y}(n+1) = y(n+1)$ can be set to any arbitrary values by $x(0)$ and $x(1)$, consider now the solutions (4) and (7) for $k = 2$.

Following similar procedure to above and using the zero-state equivalence condition, $\bar{y}(n+2) = y(n+2)$ yields

$$\bar{y}(n+2) - y(n+2) = \left[\frac{1}{\bar{a}_n(2)} - \frac{1}{a_n(2)}\right]x(2) - \left[\frac{\bar{a}_{n-1}(2)}{\bar{a}_n(2)} - \frac{a_{n-1}(2)}{a_n(2)}\right]y(n)$$
$$- \left[\frac{\bar{a}_{n-2}(2)}{\bar{a}_n(2)} - \frac{a_{n-2}(2)}{a_n(2)}\right]y(n+1) = 0$$

for all arbitrary values of $x(2)$, arbitrarily and independently set values of $y(n+1)$ and $y(n)$. The validity of this equation is possible if and only if the coefficients of $x(2), y(n+1)$ and $y(n)$ are zero. Hence it is straight forward to drive

$$\bar{a}_n(2) = a_n(2), \bar{a}_{n-1}(2) = a_{n-1}(2), \bar{a}_{n-2}(2) = a_{n-2}(2). \tag{8c}$$

Continuing this way for $k = 3,4,\cdots,n$, the process yields

$$\bar{a}_n(k) = a_n(k), \bar{a}_{n-1}(k) = a_{n-1}(k), \cdots, \bar{a}_{n-k}(k) = a_{n-k}(k) \tag{8d}$$

for $k = 0,1,\cdots,n$; and for $k \geq n+1$

$$\bar{a}_n(k) = a_n(k), \bar{a}_{n-1}(k) = a_{n-1}(k), \cdots, \bar{a}_0(k) = a_0(k). \tag{9}$$

Thus combining the results in Eqs. (8) and (9), we arrive the result in Eq. (6). Hence the lemma is proved. So the lemma is proved.

Note that the equivalence of $\bar{a}_i(k)$ and $a_i(k)$ is not necessary for $k = 0,1,\cdots,n-i-1$. Since these are the coefficients coupling the initial values $\bar{y}(0), \bar{y}(1), \cdots, \bar{y}(n-1)$ and $y(0), y(1), \cdots, y(n-1)$ to $\bar{y}(k)$ and $y(k)$, respectively, and these initial conditions are zero for the zero-state response; therefore the mentioned coefficients may not be equal.

Zero-input equivalence of systems (3) and (5) can be defined similarly by considering $x(k) = \bar{x}(k) \equiv 0, \forall k \geq 0$ as follows:

**Definition 2:** Two discrete-time linear time-varying systems of the same order $n$ described by the difference equations (3) and (5) are said to be zero-input equivalent if they produce the same outputs for the same set of arbitrarily chosen initial conditions; that is

$$\bar{y}(k) = y(k), \forall k \geq 0 \tag{10a}$$

for zero-inputs

$$\bar{x}(k) = x(k) \equiv 0, \forall k \geq 0 \tag{10b}$$

and for all arbitrarily chosen initial states

$$\bar{y}_k = y_k, \quad k = 0,1,2,\cdots, n-1. \tag{10c}$$

**Lemma 2:** For the zero-input equivalence of the systems (3) and (5), it is sufficient but not necessary that one system is an algebraic multiple of the other, that is

$$\bar{a}_i(k) = \alpha_k a_i(k), i = 0,1,2,\cdots,n; \forall k \geq 0, \tag{11a}$$

where the non-zero finite constants $\alpha_k$ are given by

$$\alpha_k = \frac{\bar{a}_n(k)}{a_n(k)}, k = 0,1,2,\cdots. \tag{11b}$$

The proof of the lemma directly follows from Eqs. (4) and (7) with $x(k) = \bar{x}(k) \equiv 0$. The equivalence of these zero-input solutions with the same arbitrary set of initial states for the systems (3) and (5) for $k = 0,1,2,\cdots$ follows with the conditions in (11); thus, the sufficiency proof ends. In fact with (11a), both systems (3a) and (5a) with $\bar{x}(k) = x(k) \equiv 0$ could be made identical by multiplying all the coefficients of (3a) by $\alpha_k$, or dividing all the coefficients of (5a) by $\alpha_k, k \geq 0$.

The non-necessity could be shown by a counter example; let the two systems be defined as

$$a_1(k)y(k+1) + a_0(k)y(k) = x(k); y(0) = y_0, \tag{12a}$$

$$\bar{a}_1(k)\bar{y}(k+1) + \bar{a}_0(k)\bar{y}(k) = \bar{x}(k); \bar{y}(0) = \bar{y}_0 \tag{12b}$$

for $k \geq 0$. Assuming $x(k) = \bar{x}(k) = 0, \forall k \geq 0$ and arbitrary equal initial condition $\bar{y}_0 = y_0$, both systems yield equal solutions $\forall k \geq 0$ which are identically equal to zero if $a_0(0) = \bar{a}_0(0) = 0$. Hence, the condition (11a) is not necessary for all $k \geq 1$.

We now ready to define the equivalence of the systems (3) and (5) in general.

**Definition 3:** Two discrete-time linear time-varying systems of order $n$ described by Eqs. (3) and (5) are equivalent if they produce the same solutions for all $k \geq 0$ for all equal input functions and equal initial conditions; hence, equivalent systems have the same input-output pairs.

**Lemma 3:** For the equivalence of systems (3) and (5) it is necessary and sufficient that

$$\bar{a}_i(k) = a_i(k), \text{ for } i = 0,1,2,\cdots,n; \forall k \geq 0, \tag{13a}$$

$$\bar{y}_k = y_k, \text{ for } k = 0,1,2,\cdots, n-1, \tag{13b}$$

that is both systems are identical.

The proof of this lemma follows from the results of Lemma I and II. In fact, from the linearity, the complete solution is the summation of the zero-state and zero-input solutions, which are independently found from each other. Therefore, Lemma I requires $\bar{a}_i(k) = a_i(k)$ for $i = n, k = 0,1,2,\cdots$; which implies $\alpha_k = 1$ in (11b) $\forall k \geq 0$. This result in turn implies (13a) from (11a). It is obviously true that the condition of Lemma I is also satisfied with (13a). Eq. (13b) follows directly from the definition of the zero-input equivalence.

**Remark 1:** In spite of the fact that the condition of Lemma 2 is not necessary, the necessity of the condition of Lemma 3 is required due to the necessity condition of Lemma 1, this is an expected result because Lemma 3 reduces to Lemma 1 for the case of zero initial conditions.

### III. COMMUTATIVITY OF FIRST-ORDER SYSTEMS

In this section, the definition of commutativity and the formulation of the commutativity problem for first-order discrete-time linear time-varying systems are presented. Consider the systems $A$ and $B$ described by

$$\text{A:} \quad a_1(k)y_A(k+1) + a_0(k)y_A(k) = x_A(k); \quad y_A(0) = y_{0A}, \tag{14a}$$

$$\text{B:} \quad b_1(k)y_B(k+1) + b_0(k)y_B(k) = x_B(k); \quad y_B(0) = y_{0B} \tag{14b}$$

for $k \geq 0$; where $a_1(k) \neq 0, b_1(k) \neq 0$. When these systems are connected in cascade as shown in Fig.1 to form a single system with input $x(k)$ and output $y(k)$, the constraint equations

$$x(k) = x_A(k), \tag{15a}$$

$$y_A(k) = x_B(k), \tag{15b}$$

$$y_B(k) = y(k) \tag{15c}$$

follow. Taking Eq. (14b) for $k$ and $k+1$ and using Eq. (15b), Eq. (14a) can be written as

$$a_1(k)b_1(k+1)y(k+2) + [a_1(k)b_0(k+1) + a_0(k)b_1(k)]y_B(k+1)$$
$$+ a_0(k)b_0(k)y_B(k) = x_A(k). \tag{16}$$

Since $x_A(k) = x(k)$ and $y_B(k) = y(k)$, for the connection $AB$ the following difference equation is obtained

$$a_1(k)b_1(k+1)y(k+2) + [a_1(k)b_0(k+1) + a_0(k)b_1(k)]y(k+1)$$
$$+ a_0(k)b_0(k)y(k) = x(k) \tag{17}$$

for $k \geq 0$. The initial values $y(0)$ and $y(1)$ are obtained similarly

$$y(0) = y_B(0) = y_{0B}, \tag{18a}$$

$$y(1) = \frac{y_A(0) - b_0(0) y_B(0)}{b_1(0)} = \frac{1}{b_1(0)} y_{0A} - \frac{b_0(0)}{b_1(0)} y_{0B}. \tag{18b}$$

For the interconnection $BA$ as shown in Fig. 1(c), the constraint equations $x = x_B, y = y_A$, together with Eqs. (14) and (15) yield

$$a_1(k+1) b_1(k) y(k+2) + [a_0(k+1) b_1(k) + a_1(k) b_0(k)] y(k+1)$$
$$+ a_0(k) b_0(k) y(k) = x(k), \tag{19}$$

$$y(0) = y_A(0) = y_{0A}, \tag{20a}$$

$$y(1) = \frac{y_B(0) - a_0(0) y_A(0)}{a_1(0)} = \frac{1}{a_1(0)} y_{0B} - \frac{a_0(0)}{a_1(0)} y_{0A}. \tag{20b}$$

Naturally, Eqs. (19) and (20) could also be obtained from the corresponding equations (17, 18) by interchanging $a_i, b_i$ and $A, B$.

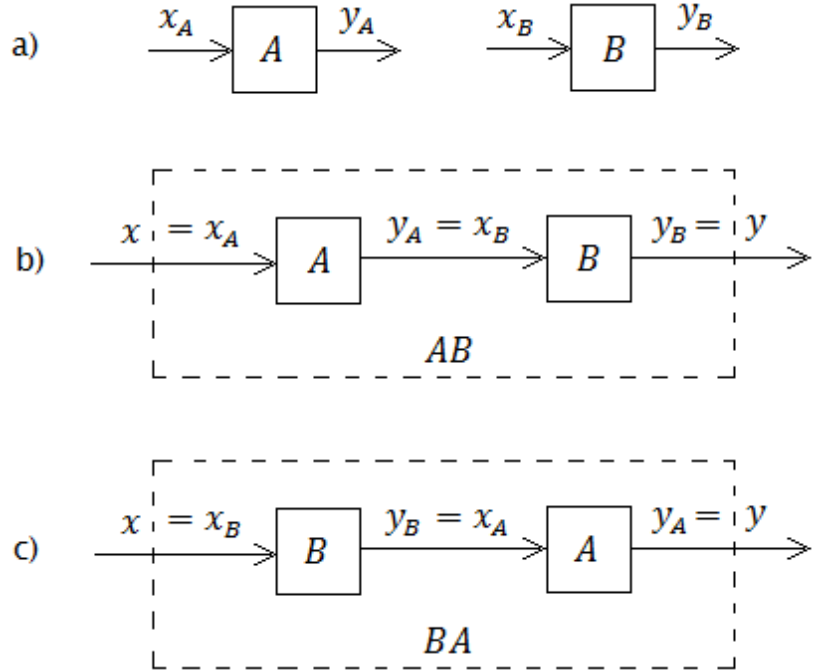

**Figure 1:** a) Two systems $A$ and $B$; and their cascade connections b) $AB$ and c) $BA$

**Definition 4:** The systems $A$ and $B$ are said to be commutative if their interconnections $AB$ and $BA$ are equivalent systems.

Since there are a few definitions for the equivalence (Definitions 1, 2, 3) considering the zero-state, zero-input and the general equivalence, respectively, the commutativity problem will be formulated accordingly.

For the commutativity of $A$ and $B$ with zero initial conditions, that is $y_{0A} = 0, y_{0B} = 0$ in Eqs. 14a and 14b, respectively, the initial conditions of the interconnection $AB$ (Eq. 18) and those of $BA$ (Eq. (20)) will be zero. Hence, the zero-state equivalence of the systems $AB$ and $BA$ is of the case. The coefficient conditions of Lemma I state the necessary and sufficient conditions as

$$a_1(k)b_1(k+1) = b_1(k)a_1(k+1); \quad k \geq 0, \tag{21a}$$
$$a_1(k)b_0(k+1) + a_0(k)b_1(k) = b_1(k)a_0(k+1) + b_0(k)a_1(k); \quad k \geq 1, \tag{21b}$$
$$a_0(k)b_0(k) = b_0(k)a_0(k); \quad k \geq 2. \tag{21c}$$

When the commutativity of $A$ and $B$ is considered due to the initial conditions only, that is without any input to the interconnections $AB$ and $BA$, the zero-input equivalence of $AB$ and $BA$ is of concern. Hence, the sufficient condition of Lemma II sets

$$b_0(k)a_0(k) = \alpha_k a_0(k)b_0(k), \tag{22a}$$
$$b_1(k)a_0(k+1) + b_0(k)a_1(k) = \alpha_k[a_1(k)b_0(k+1) + a_0(k)b_1(k)], \tag{22b}$$
$$b_1(k)a_1(k+1) = \alpha_k b_1(k+1)a_1(k), \tag{22c}$$
$$\alpha_k = \frac{b_1(k)a_1(k+1)}{a_1(k)b_1(k+1)}; \quad \forall k \geq 0. \tag{22d}$$

For the commutativity of $A$ and $B$ under general conditions, the systems $AB$ and $BA$ are required to be equivalent; hence, Lemma III sets

$$a_1(k)b_1(k+1) = b_1(k)a_1(k+1), \tag{23a}$$
$$a_1(k)b_0(k+1) + a_0(k)b_1(k) = b_1(k)a_0(k+1) + b_0(k)a_1(k), \tag{23b}$$
$$a_0(k)b_0(k) = b_0(k)a_0(k), \tag{23c}$$

for all $k \geq 0$. Furthermore, the equality of the initial conditions requires

$$y_{0B} = y_{0A}, \tag{24a}$$
$$\frac{1}{b_1(0)} y_{0A} - \frac{b_0(0)}{b_1(0)} y_{0B} = \frac{1}{a_1(0)} y_{0B} - \frac{a_0(0)}{a_1(0)} y_{AO}. \tag{24b}$$

## IV. ALTERNATE COMMUTATIVITY CONDITIONS

In the light of the definitions of commutativity and the formulation of the various commutativity problems considered in the previous sections, the alternate commutativity conditions and some special forms of commutativity are considered in this section. The new conditions which are equivalent to the explicit conditions obtained in the previous Sections are more useful to obtain all the commutative pairs ($B$) of a given system $A$. Further, they are expressed in the matrix form as in the commutativity conditions for the analog systems of any order [6]. Note that the commutativity conditions for the second-order discrete-time systems can not be written in the matrix form [7]. Moreover, the new conditions set explicitly the relation between the arbitrary constants ($c_1, c_2$) used in matrix form and $a_0(k)$ coefficient of $A$, they are also favorable to prove Theorems I and II and the related Corollaries of this Section.

Let the purpose be to find the commutative pairs of the system $A$. In this respect, (21c) is an identity and always satisfied. (21a) implies

$$b_1(k+1) = \frac{a_1(k+1)}{a_1(k)} b_1(k), \quad \forall k \geq 0. \tag{25a}$$

The solution of this difference equation for $b_1(k)$ is simply

$$b_1(k) = \frac{a_1(k)}{a_1(0)} b_1(0), \quad \forall k \geq 0. \tag{25b}$$

With this solution for $b_1(k)$, (21b) yields the following difference equation for $b_0(k)$;

$$b_0(k+1) = \frac{b_1(0)}{a_1(0)} [a_0(k+1) - a_0(k)] + b_0(k), \quad \forall k \geq 0. \tag{26a}$$

The solution of this first-order difference equation for $b_0(k)$ is

$$b_0(k) = \frac{b_1(0)}{a_1(0)} [a_0(k) - a_0(0)] + b_0(0), \forall k \geq 0. \tag{26b}$$

Since $b_0(0)$ and $b_1(0)$ can be chosen arbitrarily except $b_1(0) \neq 0$, assigning $b_0(0) = c_0$, and $\frac{b_1(0)}{a_1(0)} = c_1 \neq 0$ as arbitrary constants, (25b) and (26b) can be written as

$$\begin{bmatrix} b_1(k) \\ b_0(k) \end{bmatrix} = \begin{bmatrix} a_1(k) & 0 \\ a_0(k) - a_0(0) & 1 \end{bmatrix} \begin{bmatrix} c_1 \\ c_0 \end{bmatrix}, \forall k \geq 0. \tag{27}$$

where $c_0$ and $c_1$ are arbitrary constants except $c_1 \neq 0$.

When the initial conditions are zero, Eqs. (24a) and (24b) are satisfied. However, with nonzero initial conditions $y_{0B} = y_{OA} \neq 0$, Eq. (24b) requires

$$c_0 = 1 - c_1 + c_1 a_0(0). \tag{28}$$

The results that have been obtained so far in this section can be expressed by a theorem.

**Theorem I (Commutativity of first-order systems):** For the commutativity of first-order discrete-time linear time-varying systems described by Eqs. (14a) and (14b), it is necessary and sufficient that the coefficients of $B$ are expressed in terms of those in $A$ as in Eq. (27) where $c_0$ and $c_1 \neq 0$ are arbitrary constants and both systems have equal initial conditions $y_{0B} = y_{0A}$; furthermore,

$c_0 = 1 - c_1 + c_1 a_0(0)$ in the case of nonzero initial conditions.

A few results of the above theorem are stated as corollaries.

**Corollary 1:** Any first-order commutative pair of a first-order discrete-time linear time-varying system is also time-varying.

The proof is apparent from Eq. (27).

**Corollary 2:** A first-order discrete-time linear time-varying system is always commutative with its pair which is feedback controlled by arbitrary constant feedback and feed forward path gains ($\beta$ and $\alpha$, respectively); it is necessary that both of the feed-gains should be constant for commutativity, that is no commutative pairs exist with variable feed gains. Conversely, all the commutative pairs of a first-order discrete-time linear time-varying system can be obtained by using the constant feed forward and feedback gains applied to it. Moreover, if the initial conditions exist, these gains should satisfy $\beta = 1 - 1/\alpha$.

**Proof:** Consider the original system $A$ and its feedback controlled version as shown in Fig. 2. It is obvious that

$$x_A(k) = \alpha[x_B(k) - \beta y_A(k)], \tag{29a}$$

$$y_A(k) = y_B(k), \tag{29b}$$

$$y_{OA} = y_{OB}. \tag{29c}$$

Inserting these equations in Eq. (14a) and arranging, we obtain

$$\frac{a_1(k)}{\alpha} y_B(k+1) + \left(\frac{a_0(k)}{\alpha} + \beta\right) y_B(k) = x_B(k), \tag{30a}$$

$$y_{OB} = y_{OA}. \tag{30b}$$

Comparing with Eq. (27), the coefficients of this system can be written as

$$\begin{bmatrix} \dfrac{a_1(k)}{\alpha} \\ \dfrac{a_0(k)}{\alpha} + \beta \end{bmatrix} = \begin{bmatrix} a_1(k) & 0 \\ a_0(k) - a_0(0) & 1 \end{bmatrix} \begin{bmatrix} \dfrac{1}{\alpha} \\ \dfrac{a_0(0)}{\alpha} + \beta \end{bmatrix}. \tag{31}$$

Hence, with the arbitrary constants

$$c_1 = \frac{1}{\alpha} \neq 0, \tag{32a}$$

$$c_0 = \frac{a_0(0)}{\alpha} + \beta, \tag{32b}$$

the conditions of Theorem I are satisfied. Furthermore, since $y_{OB} = y_{OA}$ is always the case and for nonzero initial conditions, the constraint (28) yields

$$\beta = 1 - \frac{1}{\alpha} \tag{33}$$

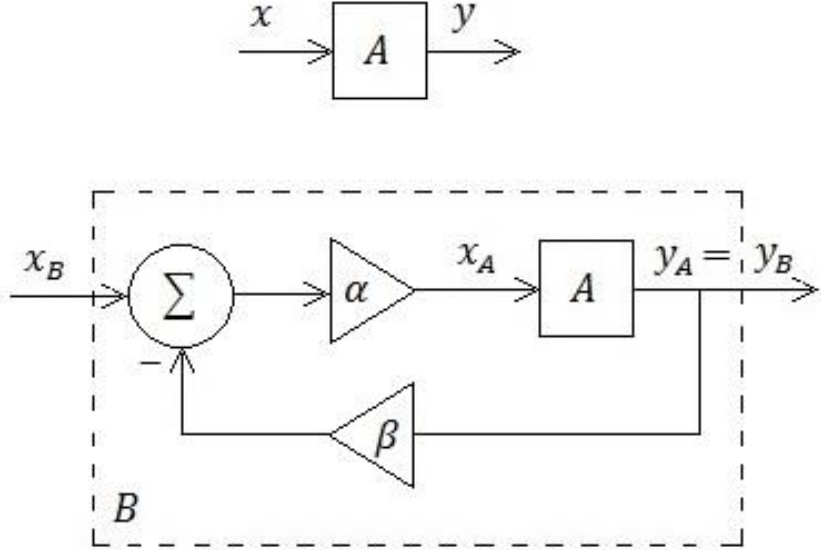

**Figure 2:** System $A$ and its feedback controlled version $B$; $\alpha \neq 0$.

on the feedback gain. Though Eqs. (29a) and (30a) are valid for variable $\alpha$ and $\beta$ as well, Eqs. (32a) and (32b) necessitate $\alpha$ and $\beta$ being constants. This completes the first part of the proof. The proof of the converse case follows from Eqs. (32a) and (32b); in fact for any $c_1 \neq 0, c_0$, the feed forward and feedback gains can be chosen as

$$\alpha = \frac{1}{c_1}, \tag{34a}$$

$$\beta = c_0 - c_1 a_0(0), \tag{34b}$$

which naturally satisfy Eq. (33) for the equal nonzero initial conditions due to Eq. (28).

We now consider commutativity conditions of a first-order system with a zero order one. First, we state the theorem and then give the proof.

**Theorem II:** A first-order discrete-time linear time-varying system is commutative with a zero-order discrete-time linear system if and only if the zero order system is time-invariant; and moreover, if the first-order system has a nonzero initial condition, then the zero order system is an identity.

**Proof:** Let the first and zero order systems denoted by $A$ and $B$, respectively and connected as in Figs. 1b and c. Let the system $A$ be represented as in (14a), and $B$ as

$$b_0(k)y_B(k) = x_B(k); \ k \geq 0, \qquad b_0 \neq 0, \tag{35a}$$

$$y_{0B} = \frac{x_B(0)}{b_0(0)}. \tag{35b}$$

For the connection $AB$, the constraints $x_A = x, y_A = x_B, y_B = y$ yield

$$a_1(k)b_0(k+1)y(k+1) + a_0(k)b_0(k)y(k) = x(k); y(0) = y_{0B} = \frac{y_A(0)}{b_0(0)}. \tag{36a}$$

For the connection $BA$, $x_B = x, y_B = x_A, y_A = y$ yield

$$a_1(k)b_0(k)y(k+1) + a_0(k)b_0(k)y(k) = x(k); y(0) = y_{0A}. \tag{36b}$$

Using the conditions of Lemma I for the equivalence $AB$ and $BA$, $a_1(k)b_0(k+1) = a_1(k)b_0(k)$ which implies

$$b_0(k+1) = b_0(k), \forall k \geq 0 \tag{37}$$

since $a_1(k) \neq 0$. Hence, the zero order system must be time-invariant. Furthermore, the equality of the initial conditions is satisfied if and only if $y_A(0) = \frac{y_A(0)}{b_0(0)}$. This means $b_0(0) = 1$; hence, if $A$ has a nonzero initial condition it has no zero-order commutative pair except identity, that is $b_0(k) \equiv 1, \forall k \geq 0$ and $y_B(k) = x_B(k), \forall k \geq 0$.

V. **EXAMPLES**

**Example 1**

This example mainly validates the theoretical results and proves the possible use of commutativity to reduce disturbance in cascade connected systems.

Let the system $A$ be described by

$$e^k y_A(k+1) + (k+1)^2 y_A(k) = x_A(k); \ y_{0A} = y_A(0) = 2, k \geq 0. \tag{38}$$

With $c_1 = 2$ and $c_0 = 1$, Eq. 27 yields the following commutative pair as system $B$:

$$2e^k y_B(k+1) + (2k^2 + 4k + 1)y_B(k) = x_B(k); \ y_{0B} = y_B(0) = 2, k \geq 0. \tag{39}$$

Note that both systems $A$ and $B$ have the same initial conditions $y_{0A} = y_{0B} = 2$ as implied by Eq. (24a) and due to nonzero value of the initial conditions $c_1$ and $c_0$ are chosen so that Eq. (28) is satisfied.

The simulation of the interconnected systems $AB$ and $BA$ is worked by Simulink for an input of unit sample sequence and seen that both systems yield the same output solution as shown in Fig. 3.

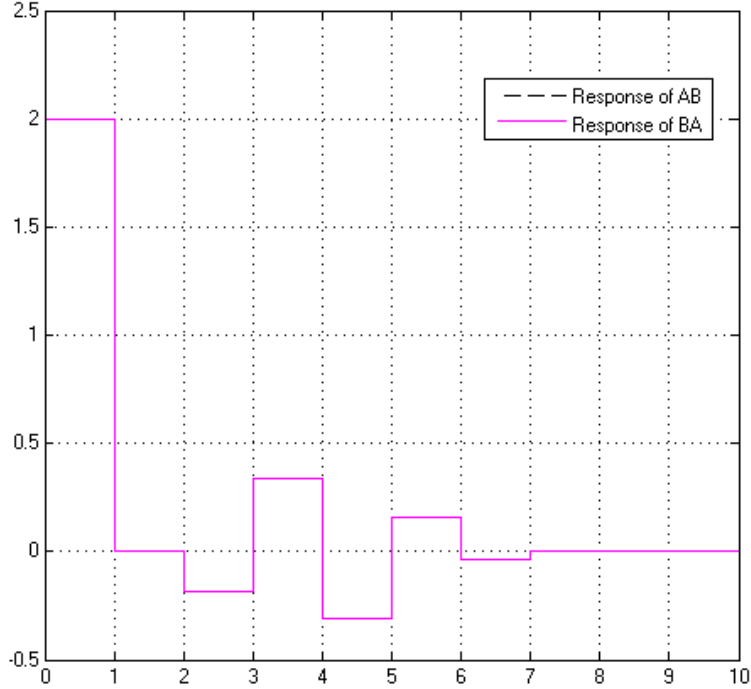

**Figure 3:** The response of the systems $AB$ and $BA$ when all the commutativity conditions are satisfied.

For the case $c_1 = 2$ and $c_0 = 3$ and with the same nonzero initial conditions, system $B$ becomes

$$2e^k y_B(k+1) + (2k^2 + 4k + 3)y_B(k) = x_B(k); \quad y_{0B} = y_B(0) = 2, k \geq 0. \quad (40)$$

For this case, although the form of $B$ is obtained from Eq. (27) and the same nonzero initial condition $y_{0A} = y_{0B} = 2$ exists, it is expected that the systems $A$ and $B$ are not commutative; this is because Eq. (28) is not satisfied. In fact, the interconnections $AB$ and $BA$ yield different output responses as shown in Fig. 4. When the initial conditions are zero, Eq. (28) is not necessary for the commutativity, therefore, both systems $AB$ and $BA$ yield the same output shown in Fig. 5 ($AB = BA$).

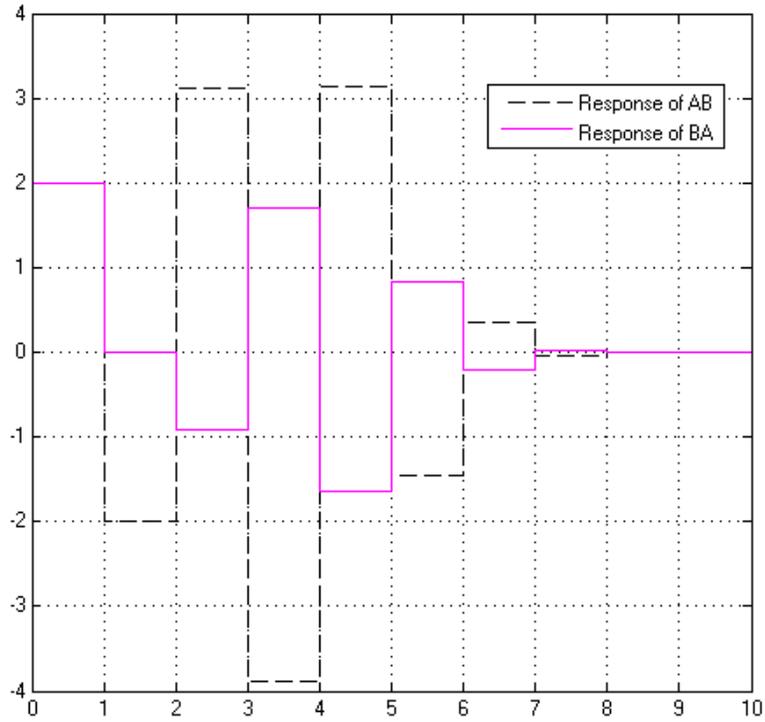

**Figure 4:** The outputs of the systems *AB* and *BA* when all the commutativity conditions except Eq. (23) are satisfied.

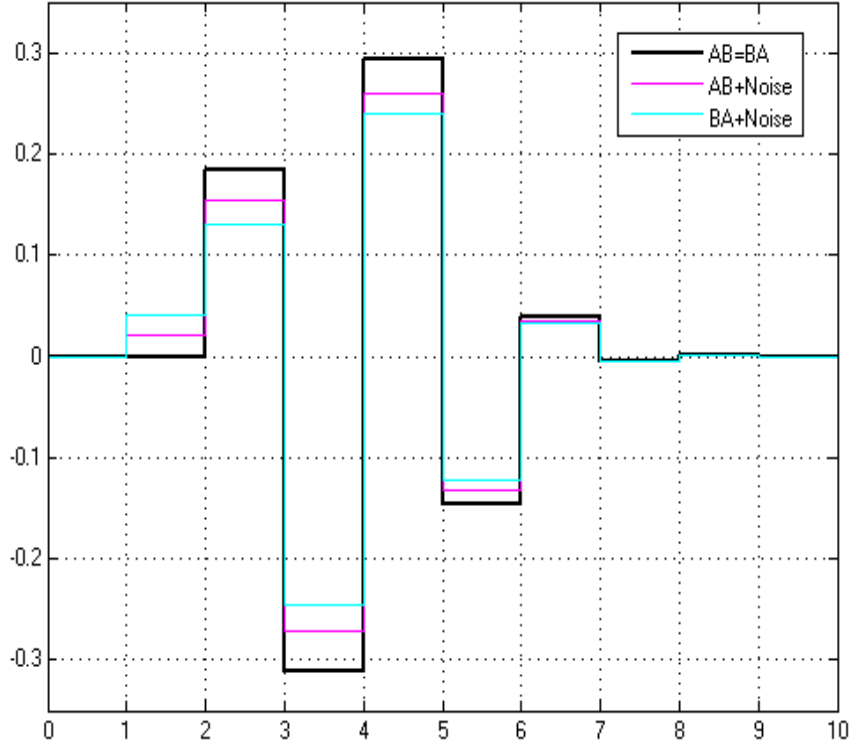

**Figure 5:** The outputs of the systems $AB$ and $BA$ without disturbance and with the same additive disturbance between $A$ and $B$; with zero initial conditions whilst Eq. (23) is still not satisfied.

To show the effect of disturbance on the system performance, a noise $0.04^{-k}$ is added between the interconnection of $A$ and $B$ for the case considered in Fig. 5 ($AB = BA$); that is $c_1 = 2$ and $c_0 = 3$ and zero initial conditions are chosen. The outputs of $AB$ and $BA$ are shown in the same figure ($AB +$ Noise and $BA +$ Noise, respectively). Comparing these curves in Fig. 5, it is seen that the interconnection $AB$ is more robust than $BA$. In fact, it is recorded that the disturbance due to noise in the response of $BA$ remains in the limits $[-0.0656, 0.0534]$ while that of $AB$ remains in $[-0.0395, 0.0324]$ which is much smaller than that of $BA$.

Note that this result is not particular for the applied disturbance $0.04^{-k}$, but it is valid for all kinds of disturbances. To prove this, consider a unit sample disturbance $\bar{\delta}(k - l)$ defined by

$$\bar{\delta}(k - l) = \begin{cases} 1, & for\ k = l \geq 0 \\ 0, & for\ k \neq l \end{cases}$$

applied to the interconnection point of $BA$. This disturbance in fact an additive input to the system $B$ defined by Eq. (40). The repetitive solution of this equation for $y_B(k)$ with $x_B(k) = \bar{\delta}(k - l)$ will be

$$h_B(k,l) = \begin{cases} 0, & \text{for } k = 0,1,\cdots,l, \\ \dfrac{1}{2e^l}, & \text{for } k = l+1, \\ \dfrac{(-1)^{k-l-1}}{2^{k-l-1}e^{(k^2-l^2+l-k)/2}} \prod_{i=1}^{k-l-1}[(k-i+1)^2+0.5], & \text{for } k \geq l+2. \end{cases} \quad (41)$$

In fact, this is the unit sample response of $B$ (unit sample stimulus occurring at $l \geq 0$). Due to linearity, the response of $B$ for any input $x(k)$ which can be expressed by

$$x(k) = \sum_{l=0}^{\infty} x(l)\,\bar{\delta}(k-l) \quad (42)$$

is given by the superposition and the result is

$$y_B(k) = \sum_{l=0}^{\infty} x(l)\,h_B(k,l) \quad (43)$$

which is known as convolution summation. Hence, the effect of the value of any $x(k)$ at any $k = l$ on $y_B(k)$ is proportional with $h_B(k,l)$.

On the other hand, for any noise $x(k)$ applied at the interconnection of $AB$ will produce

$$y_A(k) = \sum_{l=0}^{\infty} x(l)\,h_A(k,l) \quad (44)$$

where $h_A(k,l)$ is the response of $A$ to the unit sample $\bar{\delta}(k-l)$; and $h_A(k,l)$ is computed as the repetitive solution of Eq. (38) for $y_A(k)$ as

$$h_A(k,l) = \begin{cases} 0, & \text{for } k = l \geq 0, \\ \dfrac{1}{e^l}, & \text{for } k = l+1, \\ \dfrac{(-1)^{k-l-1}}{e^{(k^2-l^2+l-k)/2}} \prod_{i=1}^{k-l-1}(k-i+1)^2, & \text{for } k \geq l+2. \end{cases} \quad (45)$$

Since the effect of the value of $x(k)$ at any $k = l$ on $y_A(k)$ is proportional with $h_A(k,l)$ due to Eq. (44), it is sufficient to compare $h_B(k,l)$ and $h_A(k,l)$ to investigate the general effect of $x(k)$ on the outputs of the interconnections $BA$ and $AB$, respectively.

In fact, Eqs. (45) and (41) yield

$$\frac{h_A(k,l)}{h_B(k,l)} = \begin{cases} 2, & \text{for } k = l+1, \\ 2\prod_{i=1}^{k-l+1}\dfrac{1}{1+\dfrac{0.5}{(k-i+1)^2}}, & \text{for } k \geq l+2. \end{cases} \quad (46)$$

Obviously, the effect of $x(l)$ on the output of $BA$ will be twice its effect on the output $AB$ for $k = l + 1$. For all $k \geq l + 2$ it is true that Eq. (46) yields $h_A(k, l) > h_B(k, l)$. Hence, due to Eqs. (43) and (44) the individual value of the noise $x(k)$ at $l$ is affecting the output of system $A$ more than that of $B$ for all instants grater than $l$. Therefore, considering the overall effects of $x(k)$ for $k \geq l$ and the summations in Eqs. (43) and (44), the connection $AB$ where the output is taken from $B$ is more robust than $BA$ for any sequence of $x(k)$ applied at the interconnection. That is, this conclusion is general for the given example and it is not due to the particularly chosen noise $0.04^{-k}$.

As the final simulation to validate Corollary 2, let the system A in Fig. 2 be defined by Eq. (38) with the forward and backward feedback gains $\alpha = 2, \beta = 0.5$ which is a choice satisfying Eq. (33). Hence, the system $A$ and its feedback connected version $B$ defined in Fig. 2 are supposed to be commutative with nonzero initial conditions as well. For a step input and with the initial condition 2, the simulation results shown in Fig. 6 ($AB = BA$) confirm this fact. On the contrary if $\beta$ is changed to 1 whilst $\alpha = 2$, which is a case Eq. (33) is not satisfied, the commutativity is spoiled as shown in Fig. 6 ($AB, BA$).

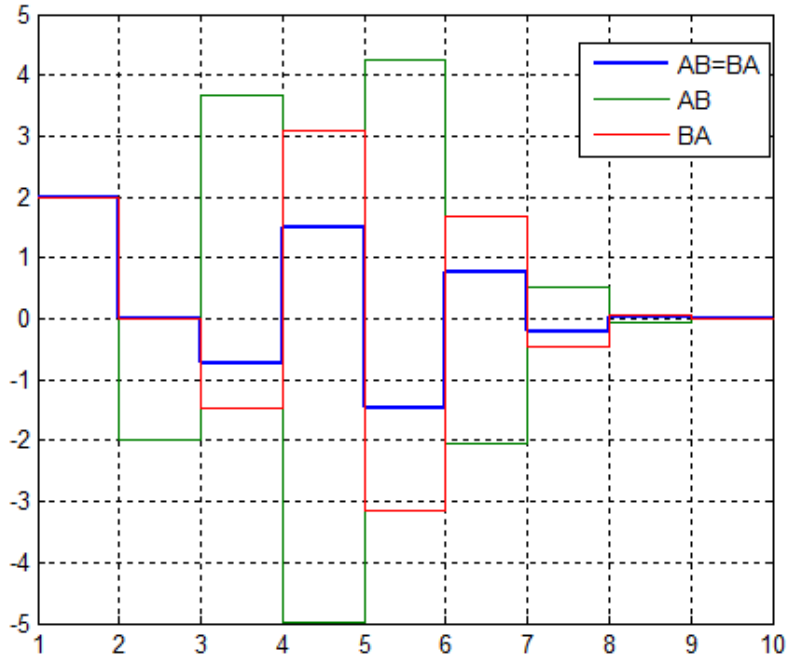

**Figure 6:** The responses of the cascade connection of a second-order system $A$ by its feedback connection $B$ with and without satisfied commutativity conditions.

**Example 2**

In this example, the commutativity concepts are illustrated by a second order low pass filter which is used as an amplitude modulator in telecommunication circuits [39]. Let the subsystems $A$ and $B$ are defined by

$$A: \quad 10y_A(k+1) + [9 + 3\sin(0.1\pi k)]y_A(k) = x_A(k), \qquad (47a)$$

$$B: \quad 30y_B(k+1) + [8 + 9\sin(0.1\pi k)]y_B(k) = x_B(k). \qquad (47b)$$

The coefficients of $A$ and $B$ satisfy Eq. 27 with $c_1 = 3$ and $c_0 = 8$; hence, $A$ and $B$ are commutative. $AB$ and $BA$ are equivalent systems which perform an amplitude modulation with a carrier frequency of 0.05. In fact, with a sinusoidal input of amplitude 100 and frequency of 0.0025, the typical output of the modulators $AB$ and $BA$ is shown in Fig. 7 after adding the combination of carrier and input signals $0.2\sin(0.1\pi k) - 0.155\sin(0.005\pi k)$ to obtain a modulation index of 27.9 %.

To test the disturbance effects of the noise to the modulators $AB$ and $BA$, a 50 % pulse width pulse sequence with amplitude 0.1 and period 2 is added at the interconnection between $A$ and $B$. The output of the modulators $AB$ and $BA$ with and without noise are shown in Fig. 8. To see the effects of noise better, the outputs are redrawn in Fig. 9 for the time interval $[400\ 440]$. It is obvious from this figure that the modulator $AB$ is less sensitive to noise interfered than the modulator $BA$. Hence, although $AB$ and $BA$ give the same modulated signals shown in Fig. 7-9 in the ideal case because $A$ and $B$ are commutative, the sequence of connection $AB$ is much better in performance when an interference is present at the interconnection.

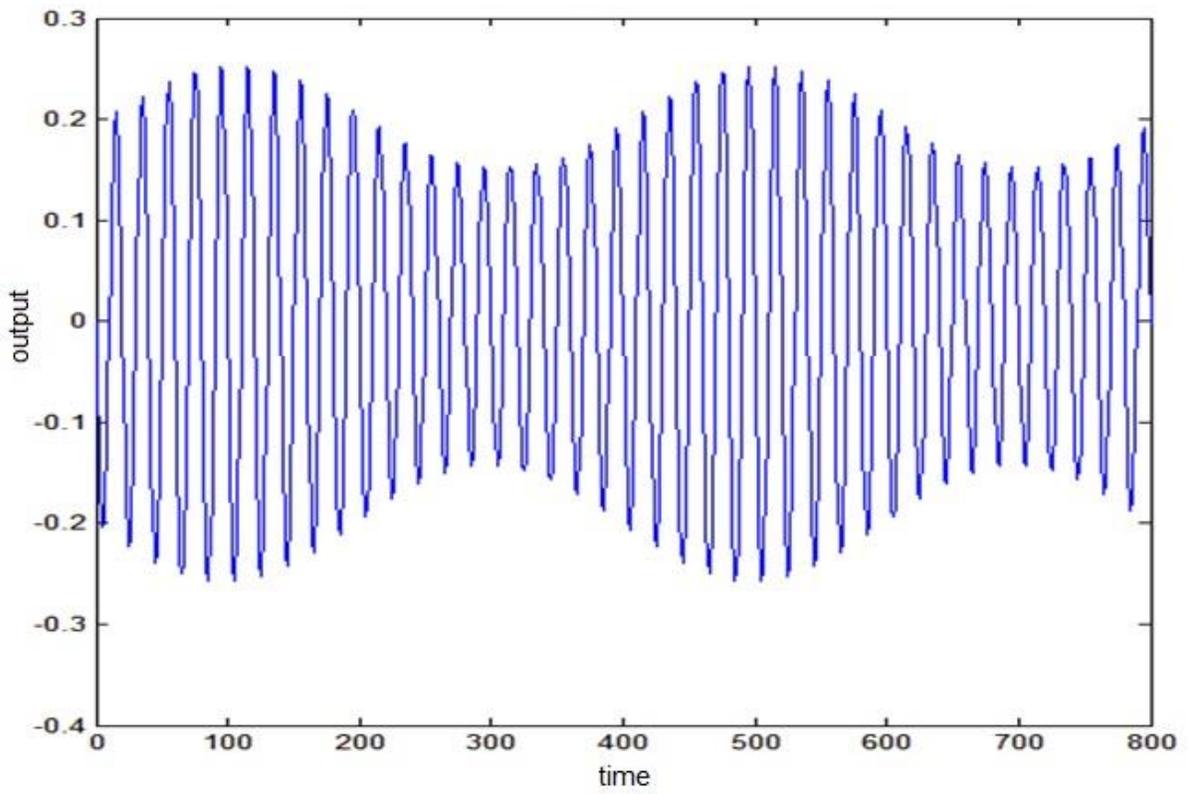

**Figure 7:** Output of modulators of Example 2 without distortion.

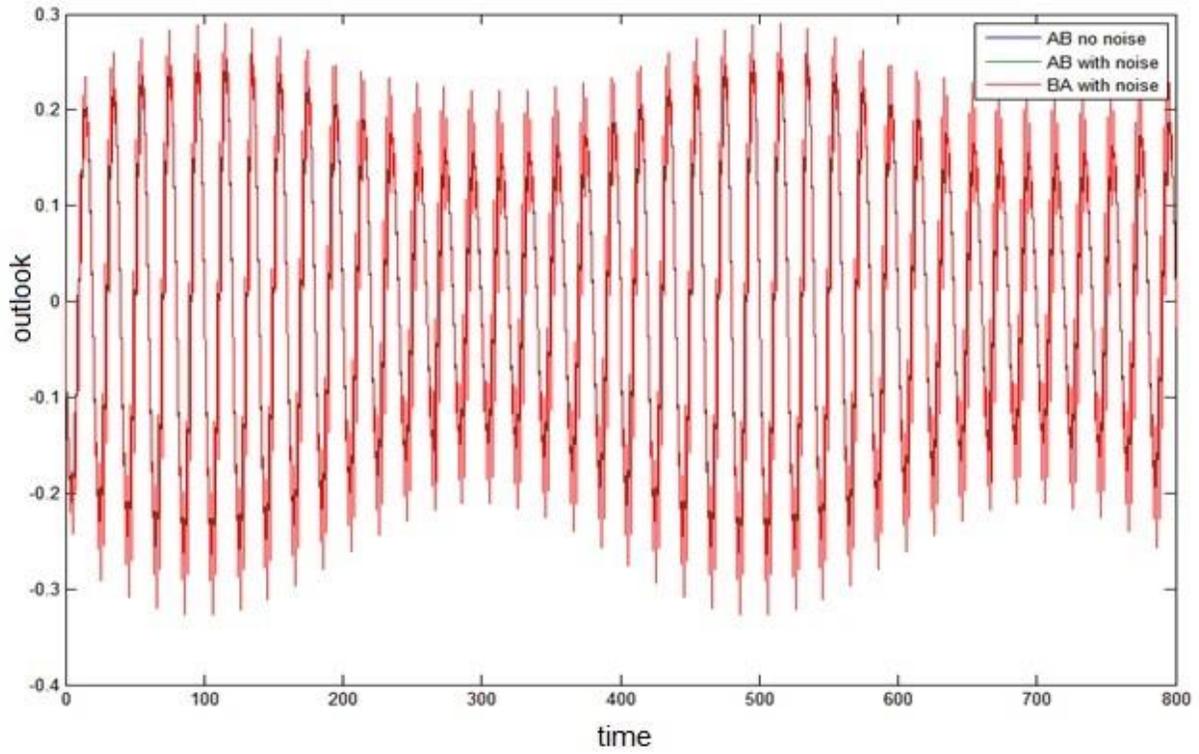

**Figure 8:** Output of modulators of Example 2 with and without distortion for time [0, 800].

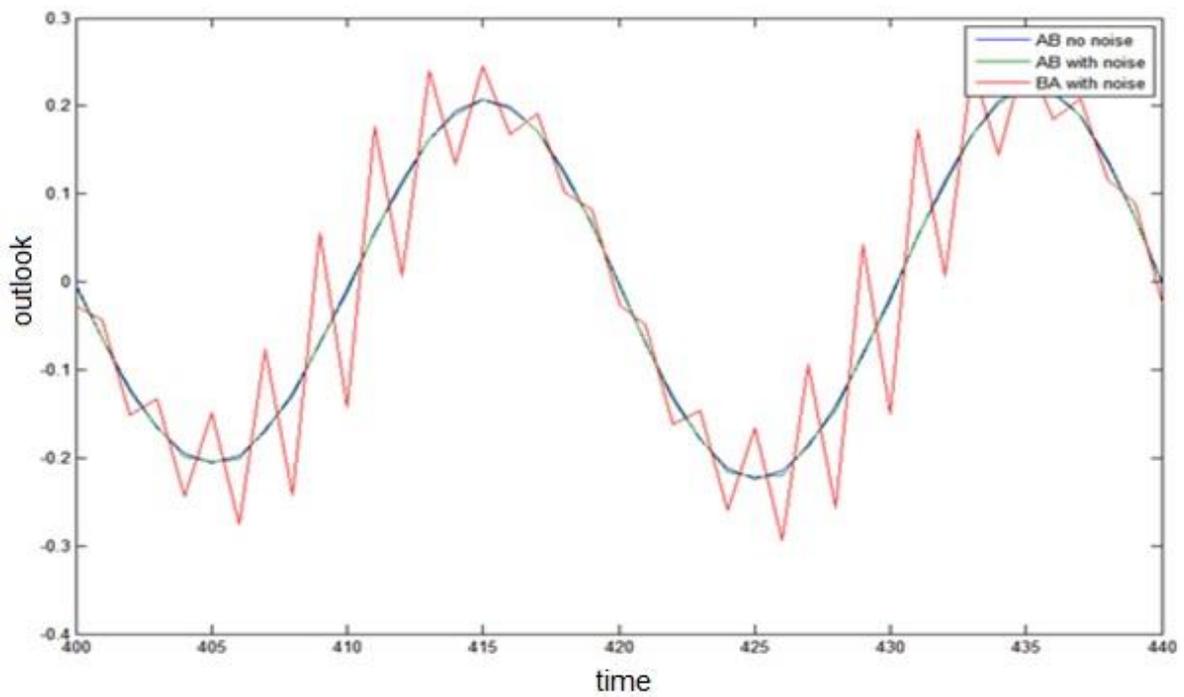

**Figure 9:** Output of modulators of Example 2 with and without distortion for time [400, 420].

## VI. CONCLUSIONS

After introducing the commutativity and commutativity conditions about the continuous-time linear time-varying systems, the commutativity concept for discrete-time systems is presented for the first time, as far as the author's knowledge, in this paper. Explicit commutativity conditions are derived and proven for the first-order discrete-time linear time-varying systems. The results are presented in a format similar to those of continuous-time systems although they are quite different from them. In this respect, the validity of the commutativity of a system with its feedback-controlled version is also verified.

An example is given to show the validity of the results and the possible use of the commutativity in applications for decreasing the disturbance effects. Many other applications are considered in various digital systems [7, 35-38].

Future work for the commutativity of higher order systems and further advantages of the commutativity properties would provide original results on the subject.

**REFERENCES**

1. A.G.J. Holt and K.M. Reineck, "Transfer function synthesis for a cascade connection network," IEEE Transactions on Circuit Theory, vol. 15, no. 2, pp. 162-163, 1968.
2. M.R. Ainbund and I.P. Maslenkov, "Improving the characteristics of microchannel plates in cascade connection," Instruments and Experimental Techniques, vol. 26, no. 3, pp. 650-652, 1983.
3. I. Gohberg, M.A. Kaashoek, and A.C.M. Ran, "Partial role and zero displacement by cascade connection," SIAM Journal on Matrix Analysis and Applications, vol. 10, no. 3, pp. 316-325, 1989.
4. B.T. Polyak, and A.N. Vishnyakov, "Multiplying disks: robust stability of a cascade connection, European Journal of Control," vol. 2, no. 2, pp. 101-111, 1996.
5. J. Walczak, and A. Piwowar, "Cascade connection of a parametric sections and its properties," Przeglad Elektrotechniczny, vol. 86, no. 1, pp. 56-58, 2010.

6. M. Koksal and M. E. Koksal, "Commutativity of linear time-varying differential systems with non-zero initial conditions: A review and some new extensions," Mathematical Problems in Engineering, vol. 2011, pp. 1-25, 2011.
7. M. Koksal and M. E. Koksal, "Commutativity of cascade connected discrete-time linear time-varying systems," Transaction of the Institute of Measurement and Control, vol. 37, no. 5, pp. 615-622, 2015.
8. R.L. Boylestad and L. Nashelsky, "Electronic Devices and Circuit Theory," Phipe Prentice Hall, 2002.
9. R.C. Dorf and J.A. Svadova, "Introduction to Electric Circuits," Wiley International Edition, 2004.
10. G.M. Miller and J.S. Beasley, "Modern Electronic Communication," Prentice Hall, 2002.
11. P.H. Young, "Electronic Communication Techniques," Engle Wood Cliffs, 1994.
12. B. Skalar, "Digital Communications: Fundamentals and Applications," Prentice Hall, 2001.
13. S. Hykin, "Digital Communication," Wiley International Edition, 1988.
14. Y. Ebihara, D. Peaucelle and D. Arzelier, "Periodically time-varying memory state-feedback controller synthesis for discrete-time linear systems," Automatica, vol. 47, pp. 14-25, 2011.
15. M. Zhong, S.X. Ding, E.L. Ding, "Optimal fault detection for linear discrete time-varying systems," Automatica, vol. 46, pp. 1395-1400, 2010.
16. R.C.L.F. Oliveira and P.L.D. Peres, "Time-varying discrete-time linear systems with bounded rates of variation: stability analysis and control design," Automatica, vol. 45, pp. 2620-2626, 2009.
17. Y. Li, S. Liu, Z. Wang, "Fault detection for linear discrete time-varying systems with intermittent observations and quantization errors," Asian Journal of Control, vol. 18, no. 1, pp. 377–389, 2016.
18. L.G.V. Willigenburg and W.L.D. Koning, "Temporal stabilizability and compensatability of time-varying linear discrete-time systems with white stochastic parameters," European Journal of Control, vol. 23, pp. 36–47, 2015.
19. Z. Zhang, Z. Zhang, H. Zhang, B. Zheng and H.R. Karimi, "Finite-time stability analysis and stabilization for linear discrete-time system with time-varying delay," Journal of the Franklin Institute, vol. 351, pp. 3457-3476, 2014.

20. H.N. Nguyen, S. Olaru, P.O. Gutman and Y.M. Hovd, "Constrained control of uncertain, Time-varying linear discrete-time systems subject to bounded disturbances," IEEE Transactions on Automatic Control, vol. 60, no. 3, pp. 831-836, 2015.

21. B. Shen, S.X. Ding and Z. Wang, "Finite-horizon H∞ fault estimation for uncertain linear discrete time-varying systems with known inputs," IEEE Transactions on Circuits and Systems, vol. 60, no. 12, pp. 902-906, 2013.

22. S.M. Tabatabaeipour, "Active fault detection and isolation of discrete-time linear time-varying systems: a set-membership approach," International Journal of Systems Science, vol. 1, pp. 1-18, 2013.

23. Y. Li and M. Zhong, "Fault detection filter design for linear discrete time-varying systems with multiplicative noise," Journal of Systems Engineering and Electronics, vol. 22, no. 6, pp. 982–990, 2011.

24. Y. Lu and X. Xu, "The stabilization problem for discrete time-varying linear systems," Systems and Control Letters, vol. 57, pp. 936–939, 2008.

25. M. Arnoldy, N. Begunz, P. Gurevichx, E. Kwamey, H. Lamba and D. Rachinskii, "Dynamics of discrete time systems with a hysteresis stop operator," SIAM Journal on Applied Dynamical Systems, vol. 16, no. 1, pp. 91-119, 2017.

26. H.A. Levine and Y.J. HA, "Discrete dynamical systems in multiple target and alternate SELEX," SIAM Journal on Applied Dynamical Systems, vol. 14, no. 2, pp. 1048–1101, 2015.

27. R.A. Borges, R.C.L.F. Oliveira, C.T. Abdallah, P.L.D. Peres, "$H_\infty$ filtering for discrete-time linear systems with bounded time-varying parameters," Signal Processing, vol. 90, pp. 282–291, 2010.

28. C. A. Desoer, "Notes for a Second Course on Linear Systems," Van Nostrand Rheinhold, New York, 1970.

29. E. Marshall, "Commutativity of time varying systems," Electro Letters, vol. 18, pp. 539-540, 1977.

30. M. Koksal, "Commutativity of second-order time-varying systems," International Journal of Control, vol. 3, pp. 541-544, 1982.

31. M. Koksal, "Corrections on 'Commutativity of second-order time-varying systems'," International Journal of Control, vol. 1, pp. 273-274, 1983.
27

32. S. V. Saleh, "Comments on 'Commutativity of second-order time-varying systems'," International Journal of Control, vol. 37, pp. 1195, 1983.
33. M. Koksal, "A Survey on the Commutativity of Time-Varying Systems", METU, Technical Report no: GEEE CAS-85/1, 1985.
34. M. Koksal, "An exhaustive study on the commutativity of time varying systems," International Journal of Control, vol. 5, pp. 1521-1537, 1988.
35. E. Avci and D. Avci, "The performance comparison of discrete wavelet neural network and discrete wavelet adaptive network based fuzzy inference system for digital modulation recognition," Expert Systems with Applications, vol. 35, no. 1-2, pp. 90-101, 2008.
36. D. Jiao, J. Kim and J. He, "Efficient full-wave characterization of discrete high-density multiterminal decoupling capacitors for high-speed digital systems," IEEE Transactions on Advanced Packaging, vol. 31, no. 1, pp. 154-162, 2008.
37. A. Klein and Y. Tsividis, "Externally linear discrete-time systems with application to instantaneously companding digital signal processors," IEEE Transactions on Circuits and Systems I, vol. 58, no. 11, pp. 2718-2728, 2011.
38. F. Jin, G. Zhao and Q. Liu, "Networked control of discrete-time linear systems over lossy digital communication channels," International Journal of Systems Science, vol. 22, no. 12, pp. 2328-2337, 2013.
39. U.A. Bakshi and A.P. Godse, Analog Communication, Technical Publications, ISBN 8184311400, 9788184311402, 2010.